%% file: curvsense_main.tex
\documentclass[notitlepage,pre,longbibliography,twocolumn]{revtex4-1}
\usepackage[utf8x]{inputenc}
\usepackage{inputenc}
\usepackage{graphicx}  
\usepackage{bm}        
\usepackage{amsmath}   
\usepackage{amssymb}   
\usepackage{amsmath}
\usepackage{amssymb}
\usepackage{siunitx} 
\usepackage{upgreek}
\usepackage{longtable}
\usepackage{array}
\usepackage{MLmem}

\usepackage{xr}
\date{\today}
\begin{document}

\title{Structural symmetry and membrane curvature sensing}

\author{Federico Elías-Wolff$\,$}
\email[]{federico.elias.wolff@dbb.su.se}
\affiliation{Department of Biochemistry and Biophysics, Stockholm University, Sweden}
\author{Alexander Lyubartsev$\,$}
\email[]{alexander.lyubartsev@mmk.su.se}
\author{Erik G. Brandt$\,$}
\email[]{erik.brandt@mmk.su.se}
\affiliation{Department of Materials and Environmental Chemistry, Stockholm University, Sweden}

\author{Martin Lindén$\,$}
\email[]{bmelinden@gmail.com}
\affiliation{Department of Cell and Molecular Biology, Uppsala University, Sweden\\ Present address: Scania CV AB, Södertälje, Sweden}

\begin{abstract}
\input{abstract.ttt}  
\end{abstract}

\maketitle
%
%
\input{maintext.ttt}

\input{acknowledgements.ttt}

\renewcommand{\thesection}{S\arabic{section}}  
\renewcommand{\thetable}{S\arabic{table}}  
\renewcommand{\thefigure}{S\arabic{figure}}
\renewcommand{\theequation}{S\arabic{equation}}
\input{curvsense_main.bbl}
\clearpage
\appendix
\setcounter{section}{0}    
\setcounter{figure}{0}    
\setcounter{equation}{0}
\setcounter{table}{0}

\input{methods.ttt}

\end{document}

%% file: abstract.ttt
Symmetry is closely intertwined with the function, genetics, and chemical properties of multiprotein complexes.  Here, we explore the relation between structural symmetry and the ability of membrane proteins to sense and induce membrane curvature, which is a key factor for modulating the shape and organization of cell membranes. Using coarse-grained simulations and elasticity theory, we show that the potential for direction-dependent membrane curvature sensing is limited to asymmetric proteins, dimers, and tetramers, and argue that one should expect this anisotropy to be strongest for dimers. Odd and higher-order symmetries strongly suppress directional curvature sensing. This classification gives a new perspective on the structure-function relation for membrane proteins, and simplifies the task of translating between molecular sensing mechanisms and their large-scale cellular consequences.

%% file: maintext.ttt
Cell membranes are subject to large and dynamic deformations, mainly
driven by the ability of membrane proteins to sense, influence, and
respond to local membrane
curvature~\cite{shibata2009,mcmahon2005,baumgart2011,tonnesen2014}.
There is a growing list of molecular mechanisms of curvature sensing,
including insertion of amphipathic peptides into the lipid headgroup
region (wedge mechanism)~\cite{ford2002,cui2011}, curved protein-lipid
interfaces of peripheral membrane proteins such as BAR domains
(scaffold mechanism)~\cite{peter2004}, and non-cylindrical
transmembrane domains of integral membrane
proteins~\cite{aimon2014,tonnesen2014}. These mechanisms clearly
depend on the protein structure, and many membrane proteins are highly
symmetric \cite{goodsell2000}, but the role of protein structural
symmetry for curvature sensing has, to the best of our knowledge, not
been systematically explored. Here, we set out to do that.

The main mechanism behind protein structural symmetry is
homo-oligomerization, where multiple identical subunits form a
functional protein unit. Since reflection and inversion symmetries are
ruled out by the chirality of the $\alpha$-carbon in polypeptide
chains, and biological membranes usually have asymmetric leaflets
(different inside and outside), by far the most common symmetries for
membrane proteins are cyclic groups that describe finite-size
rotations around the membrane normal~\cite{goodsell2000}.

While many curvature sensing mechanisms depend on molecular details,
their biological consequences in terms of membrane shape and
organization typically play out on the much larger length scales of
organelles (\SI{10}{\nm} and up) or whole cells
~\cite{shibata2009,mcmahon2005,baumgart2011}. Bridging this gap of
scales is a key challenge in understanding the biological functions of
membrane curvature sensing.

\begin{figure}[t!]
  \includegraphics[width=80mm]{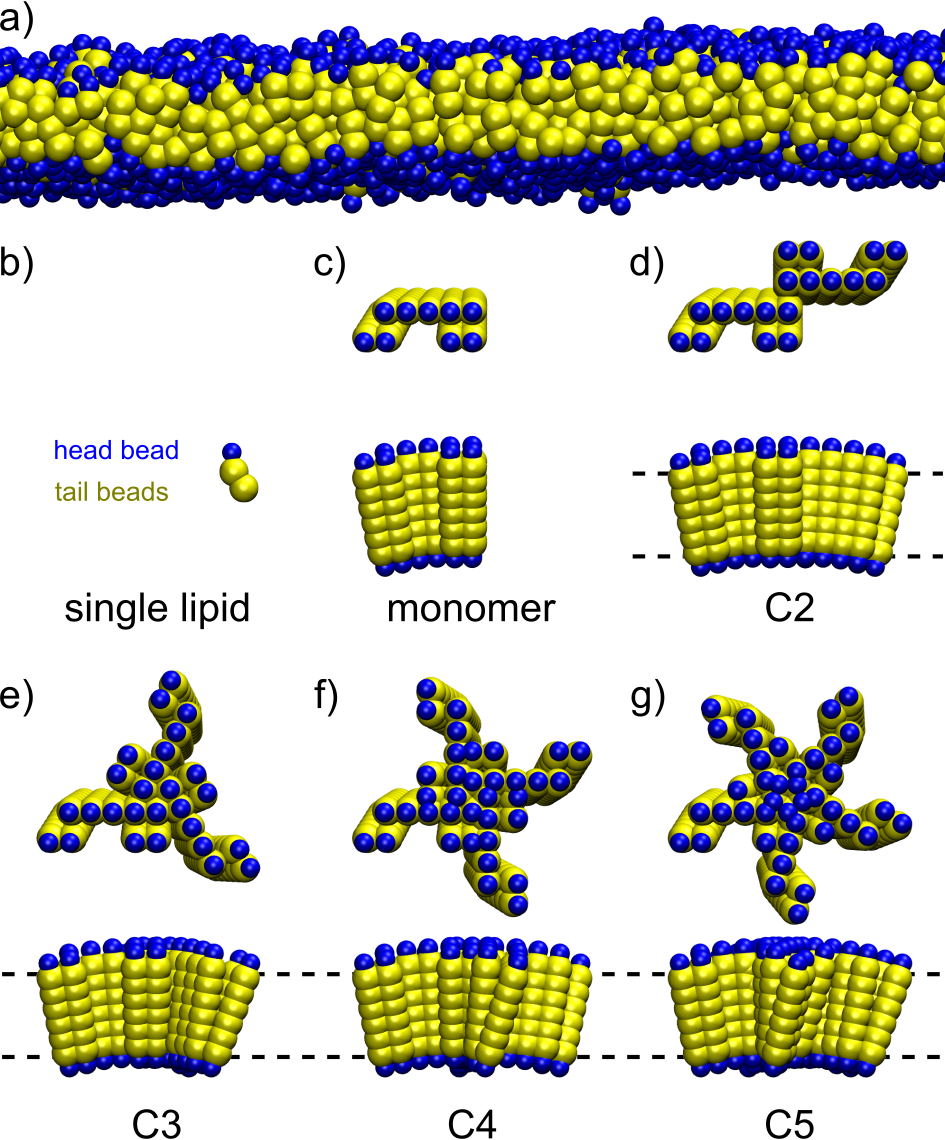}
  \caption{\label{fig:cgmodels} Coarse-grained lipids and proteins. (a)
    Side-view of a flat patch of a coarse-grained fluid membrane, which in our
    parameter regime has a thickness of $\sim 5\sigma$~\cite{cooke2005}. (b)
    Single 3-bead lipid with one head bead (blue) and two tail beads
    (yellow). (c) Protein monomer. (d-g) Top- and side views of protein
    multimers with 2- to 5-fold cyclic symmetry.}
\end{figure}

Simplified
theoretical models,
provide a natural route to this end. Membranes are then
modeled as thin elastic sheets decorated with proteins whose binding
energies are functions of the local membrane curvature. Since proteins
have complex shapes, curvature sensing will be coupled to the binding
orientation as well as the protein position~\cite{gomez2016}. This is
best described by a protein binding energy that depends on the local
curvature tensor in the protein's frame of
reference~\cite{gomez2016,ramakrishnan2014,ramakrishnan2013,akabori2011,perutkova2010,fournier1996,leibler1986}.
Numerical studies indicate that such anisotropic curvature-sensors are
capable of complex, large-scale membrane remodeling
\cite{ramakrishnan2013}.
Here, we explore how these types of interactions are restricted by
protein structural symmetry, and argue that these restrictions have
biological implications: for membrane proteins, different symmetries
are suited for different functions.


We use a coarse-grained implicit-solvent model of membranes in the fluid
phase to study generic effects of cyclic symmetry on curvature sensing. Single
lipids are represented by one head bead and two tail beads~\cite{cooke2005pre}
(Fig.~\ref{fig:cgmodels}a,b). We construct an asymmetric monomer with a slight
wedge-shaped transmembrane region from tail beads sandwiched between head
beads (Fig.~\ref{fig:cgmodels}c), and assemble cyclic multimers by in-plane
rotations (Fig.~\ref{fig:cgmodels}d-g). The model proteins are stabilized by a
network of elastic springs. The trimer in this family has a preferred
curvature of about $0.035\sigma^{-1}$ in our parameter regime
\cite{elias-wolff2018} ($\sigma\approx 1\si{nm}$ being the coarse-grained length
unit~\cite{cooke2005}). For further details, see Supplementary methods.

\begin{figure}[ht!]
  \includegraphics[width=80mm]{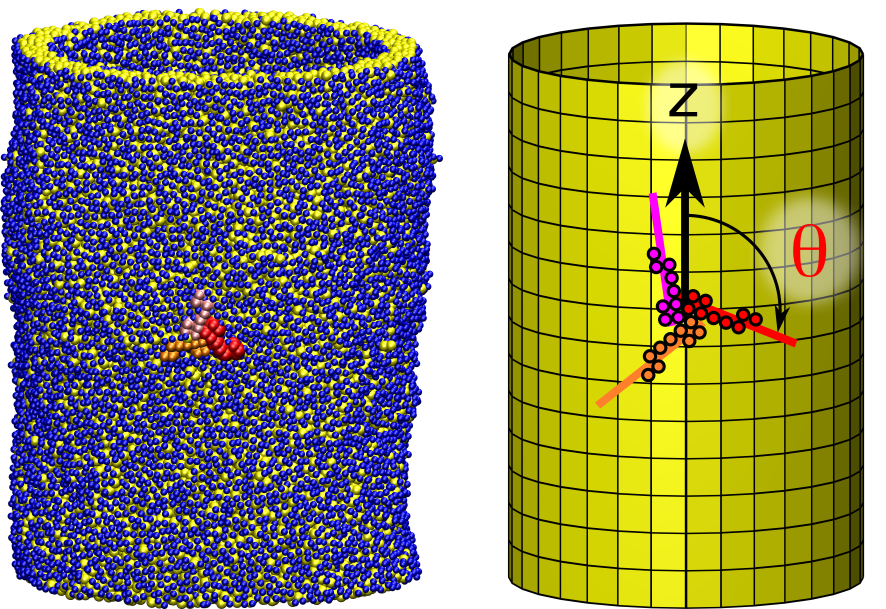}
  \caption{\label{fig:cylinder}Protein orientation on a cylinder. (Left)
    snapshot of a trimer simulation, with the monomers colored red, orange, and
    pink. (Right) Analysis sketch with in-plane angle $\theta$ for the first
    monomer.}
\end{figure}

The cyclic symmetry of the protein complex couples directly to its in-plane
orientation. We simulated proteins on a cylindrical bilayer and measured the
in-plane orientation (Fig.~\ref{fig:cylinder}). The curvature is uniform in
cylindrical geometry, so that the curvature tensor seen by the protein is
determined completely by the protein orientation, and we do not need to
constrain or keep track of the protein position.

\begin{figure}[ht!]
  \includegraphics[width=80mm]{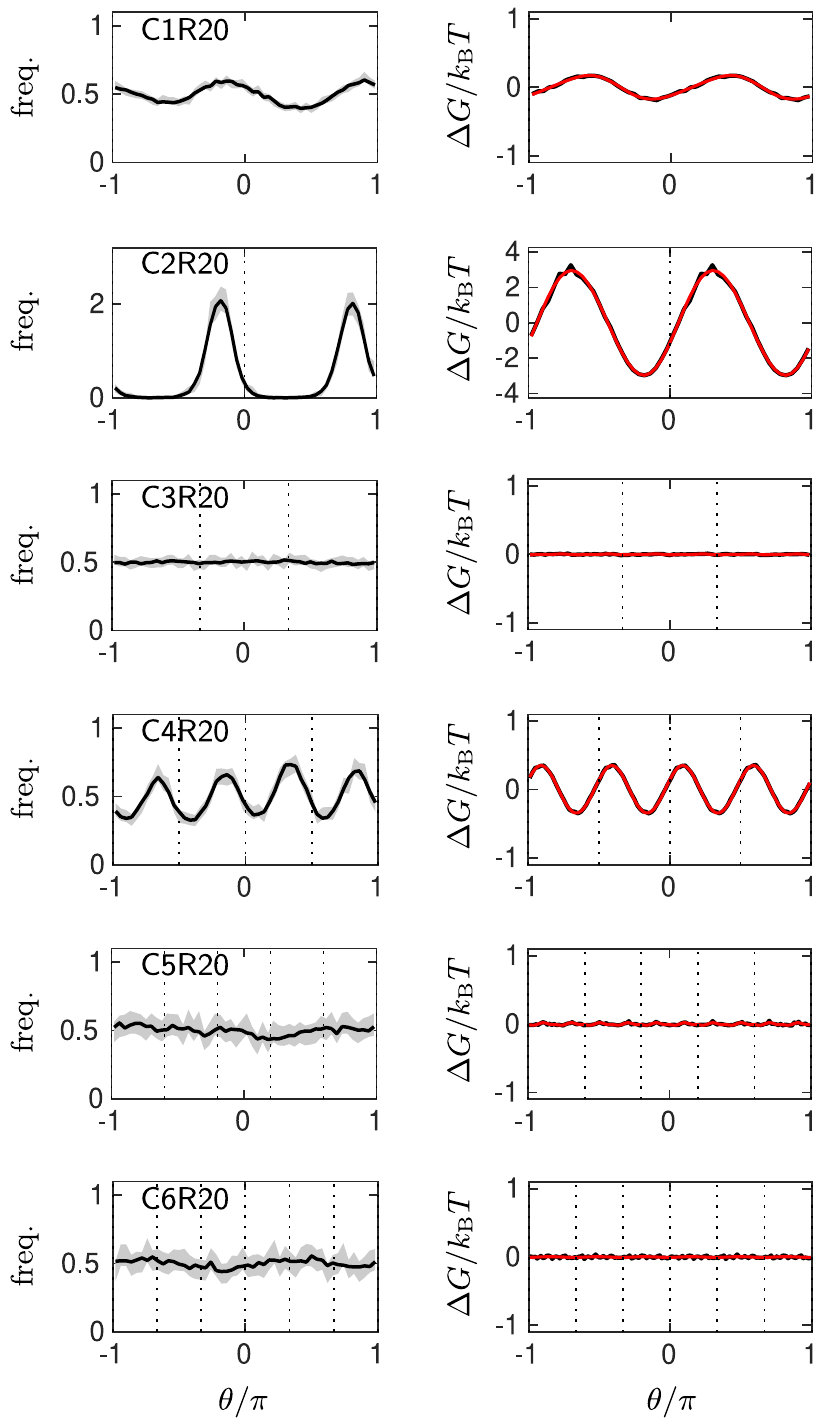}
  \caption{\label{fig:cylhist} Orientational distributions on cylinders. (Left)
    Orientational histograms of subunit one in 1- to 6-mers, with aggregated
    distributions (black) and 95\% bootstrapped confidence intervals
    (gray). (Right) Corresponding orientational free energy, where the
    symmetries of the cylinder and protein structures have been imposed to
    improve the statistics. Solid black lines are averages, and red lines are
    least-square fits to Fourier series with two terms.}
\end{figure}

\subsection*{Orientational distributions}
First, we simulated proteins with 1- to 6-fold symmetry on cylinders
with radii $R = 20\sigma$. The principal curvatures of a cylinder is
zero in the longitudinal direction, and $1/R$ in the circumferential
direction, so the local curvature experienced by the protein varies
with the orientation angle $\theta$. Fig.~\ref{fig:cylhist}(left)
shows orientational distributions of a single subunit of the different
proteins, averaged over multiple replicas with different initial
orientations (see Methods). Non-flat distributions indicate
orientational curvature sensing, and are seen in monomers, dimers, and
tetramers, with dimers showing the largest amplitude by far.  For
3-,5-, and 6-mers we see no orientational preferences within our
sampling accuracy.

The number of peaks reflect two symmetries of our setup: the cyclic
symmetry of the proteins, and invariance under rotation of the whole
system around an axis in the $(x,y)$-plane, which turns it upside down
($\theta\to\theta+\pi$). By including all subunits and enforcing the
upside down symmetry to further improve sampling, we construct the
orientational free energy profiles shown in Fig.~\ref{fig:cylhist}
(right).

\subsection*{Elastic theory}
Why do only monomers, dimers, and tetramers sense anisotropic
curvature?  A simple model can explain this behavior, if we assume
that the curvature-dependent binding free energy of a protein depends
on the local
curvature~\cite{gomez2016,ramakrishnan2014,ramakrishnan2013,akabori2011,perutkova2010,fournier1996,leibler1986}.
The curvature tensor is given by
\begin{equation}\label{eq:htensor}
  h=\left[\begin{array}{cc} h_\parallel&h_X\\h_X&h_\perp\end{array}\right]
  =\brmat{cc}{H+D\cos2\theta&D\sin2\theta\\D\sin2\theta&H-D\cos2\theta},
\end{equation}
in a basis where $\hat e_\parallel$ follows the protein orientation
and is rotated by $\theta$ w.r.t.~the principal curvature direction
$\hat e_1$.  We have introduced the mean and deviatoric curvatures
\begin{equation}
  H=(c_1+c_2)/2,\quad D=(c_1-c_2)/2,
\end{equation}
with $c_{1,2}$ being the principal curvatures. The Gaussian curvature
is $K=c_1c_2=4(H^2-D^2)$.  With $\hat e_1 = \hat e_z$ as illustrated
in Fig.~\ref{fig:cylinder}b, the principal curvatures are $c_1 = 0$
and $c_2 = 1/R$. In general, $c_{1,2}$, as well as the reference
direction for $\theta$, may vary with the protein position. We also
note that the curvature tensor is invariant under half turn rotations,
$\theta\to\theta+\pi$, just as the cylinder geometry.  In our model,
the curvature-dependent part of the binding free energy per molecule,
$G$, is a function of the local curvature, $G=G(\vec{x},\theta)
= \hat{G}(h(\vec{x},\theta))$, that can be expanded in a Taylor series
in $h_1=\frac12(h_\parallel+h_\perp)=H$,
$h_2=\frac12(h_\parallel-h_\perp)=D\cos(2\theta)$, and
$h_3=h_X=D\sin(2\theta)$,
\begin{equation}\label{eq:taylor}
  G= \sum_{i}\alpha^{(1)}_{i}h_{i} + \sum_{ij}\alpha^{(2)}_{ij}h_{i}h_{j}+\ldots,
\end{equation}
where $\alpha^{(m)}_{i,j,\ldots}$ are coefficients for terms of order $m$ in
curvature.
The lowest-order constant term is irrelevant and has been
discarded. In general, there are $(m+1)(m/2+1)$ terms of order $m$
\cite{numterms},
but symmetry can impose restrictions on the coefficients.  For
  example, combinations such as $h_\parallel+h_\perp=2H$ and
  $h_\parallel h_\perp-h_X^2=K$ are invariant under rotations and
  compatible with all structural symmetries. To see how symmetry
restricts the curvature dependence, we rewrite \Eq{eq:taylor} as a
Fourier series,
\begin{equation}\label{eq:fourier}
  G=a_0+\sum_{n=1}^\infty \left[ a_{2n} \sin(2n\theta) + b_{2n}\cos(2n\theta) \right] ,
\end{equation}
where the $a,b$ coefficients are functions of $H,D$ (or $H,K$), but
independent of $\theta$. The fact that the curvature tensor elements
in \Eq{eq:taylor} depend on $\theta$ only through
$h_2=D\cos(2\theta)$ and $h_3=D\sin(2\theta)$ means that only even
multiples of $\theta$ are included in \Eq{eq:fourier}, and that higher
order Fourier terms are also higher order in curvature. To see this,
we note that translating between the Fourier and Taylor series
expansions involves the use of trigonometric addition formulae
\cite{weisstein_multiadd}, which in this setting reads
\begin{eqnarray}
  \label{eq:sin}
  D^n\sin(2n\theta)&=&
  \sum_{k=0}^n\sin\Big[\frac\pi2(n-k)\Big]\binom{n}{k} h_2^kh_3^{n-k},\, \\
  \label{eq:cos}
  D^n\cos(2n\theta)&=&
  \sum_{k=0}^n\cos\Big[\frac\pi2(n-k)\Big]\binom{n}{k}h_2^kh_3^{n-k}.
\end{eqnarray}
It follows that the lowest-order curvature dependence in the coefficients of
$\sin(2n\theta)$ and $\cos(2n\theta)$ is $D^n$.
Moreover, the binding free energy of a protein with $M$-fold cyclic symmetry is
invariant under rotations of $1/M$ turns, i.e.,
\begin{equation}
  G(\vec{x},\theta) = G\Big(\vec{x},\theta+\frac{2\pi}{M}\Big),
\end{equation}
which is only true for those terms in Eq.~\eqref{eq:fourier} where
$2n$ is a multiple of $M$. This leaves monomers and dimers ($M=1,2$)
unrestricted, but for $M=3,4,5,6$ the lowest non-zero Fourier terms
are $2n=6,4,10,6$, which are of order $D^3,D^2,D^5,D^3$,
respectively. These are the leading-order contributions to the
orientational free energy amplitudes in Fig.~\ref{fig:cylhist}, which
suggests a simple explanation for the amplitude pattern: non-zero
linear and quadratic terms are needed to produce an angular
dependence, while the effects of higher order terms are too small to
see.  The pattern continues for higher symmetries. For even $M$-mers,
the lowest order anisotropic terms are
$D^{M/2}\cos(M\theta),\,D^{M/2}\sin(M\theta)$, while for odd $M$-mers,
the lowest order terms are
$D^M\cos(2M\theta),\,D^M\sin(2M\theta)$. Monomers, dimers, and
tetramers are therefore the only symmetries with significant
anisotropic curvature sensing at moderate curvatures. For example, the
lowest order anisotropic terms compatible with tetramers are
$D^2\sin(4\theta)=h_x(h_\parallel-h_\perp)$ and
$D^2\cos(4\theta)=\frac14(h_\parallel-h_\perp)^2-h_X^2$.

\begin{figure}[ht!]
  \includegraphics[width=80mm]{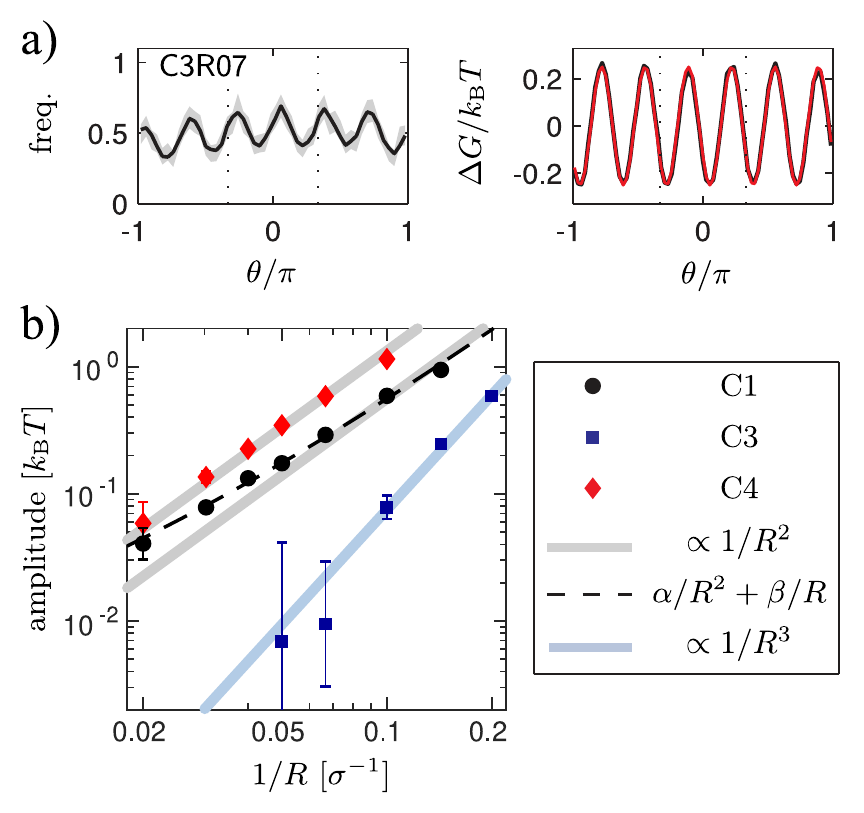}
  \caption{\label{fig:amplitude} a) Rotational distribution and free
    energy of a trimer at high curvature ($R=7\sigma$), computed as in
    Fig.~\ref{fig:cylhist}. b) Rotational free energy amplitudes
      vs $1/R$. The monomer protein (C1) amplitude is well described
      by a phenomenological fit to a general quadratic relation,
      $\alpha/R^2+\beta/R$. The trimer (C3) and tetramer (C4) show the
      expected scaling as $1/R^3$ and $1/R^2$, respectively.  Error
      bars are 95\% confidence intervals, estimated by bootstrapping
      (see Supplementary Methods).}
\end{figure}

The above theory makes testable predictions. First, higher order terms should
become prominent at high enough curvatures. In particular,
$D^3\sin(6\theta),D^3\cos(6\theta)$ should produce orientational distributions
with 6-fold periodicity for a trimer at high enough curvature. This is clearly
demonstrated when the cylinder radius is decreased from $20\sigma$ to $7\sigma$
in Fig.~\ref{fig:amplitude}a.

Second, different symmetries should produce different relationships between the
deviatoric curvature $D$ and the amplitude of the rotational free energy,
depending on the order of the lowest order terms. Since the magnitude of $D$ is
proportional to $1/R$ on a cylinder, the tetramer amplitude should
scale as $1/R^2$, while the trimer amplitude should scale as $1/R^3$. Monomers
and dimers allow anisotropic terms that are both linear and quadratic in $D$,
which implies a scaling relation involving both $1/R$ and $1/R^2$.  (All
amplitudes tend to zero at vanishing curvature.) We simulated monomers,
trimers, and tetramers on a range of cylinder radii. Fig.~\ref{fig:amplitude}b
shows the resulting amplitudes plotted against $1/R$. The results are in good
agreement with the theory: The predicted quadratic and cubic scalings for the
tetramer and trimer are clearly visible, while the monomer amplitude approaches
zero slower than $1/R^2$ for increasing $R$, as expected.

\subsection*{Discussion} Our simulations and theoretical analysis classify
membrane proteins in two categories of curvature sensing behavior
depending on their structural symmetry: Monomers and homo-multimeric
dimers and tetramers are allowed by symmetry to sense curvature
anisotropically, meaning that direction dependent terms of first or
second order are allowed in the binding free energy. Trimers,
pentamers and higher order symmetric proteins couple to curvature
isotropically, meaning that only direction-independent low-order terms
are allowed.  Our analysis also suggests that dimers are the strongest
anisotropic curvature sensors, as in \Fig{fig:cylhist}. The 2-fold
symmetry of the curvature tensor allows dimeric subunits to amplify
each other constructively, while subunits in proteins with higher
symmetry partly counteract each other by canceling out low order
terms.

These results have biological implications. Anisotropic curvature
sensing could be useful for some protein functions, but not others,
which implies a correlation between structural symmetry and the
biological function of membrane proteins. Proteins that stabilize,
localize to, and/or function on, anisotropically curved membrane
structures such as ridges, tubes, and pores, could be expected to
sense anisotropic curvature, and hence be found as monomers, dimers,
or tetramers. BAR domain proteins, that stabilize membrane
tubules~\cite{peter2004,frost2009} for example during
endocytosis~\cite{liu2009}, are indeed dimers. The (hetero-multimeric)
functional units of the rotary motor protein F$_1$F$_o$ (ATP synthase)
can dimerize in order to assemble along, and stabilize, the highly
curved ridges of mitochondrial inner membranes~\cite{davies2012}. The
M2 proton channel, which mediates virus budding and localizes to the
neck of budding virions~\cite{rossman2010}, is a tetramer. In addition
to localization and stabilization of anisotropic membrane structures,
anisotropic curvature sensing also allows the membrane shape to
organize the in-plane orientation of proteins. This could facilitate
the formation of regular protein coats, for example by BAR domains
bound to membrane tubules at high density~\cite{frost2008}.
On the other hand, membrane proteins that do not need to sense
anisotropic curvature may profit from being structurally prohibited to
do so. One example might be ion channels gated by membrane tension,
which is an isotropic mechanical signal. Notable members of this class
are the mechanosensitive channels of large (MscL,
pentamer~\cite{chang1998}) and small (MscS, heptamer~\cite{bass2002})
conductance, and the trimeric ion channel Piezo1~\cite{guo2017}.
Beyond these examples, bioinformatic methods may be able to probe
correlations between biological function and structural symmetry in
membrane proteins more systematically.

Experimentally, anisotropic curvature sensors are difficult to characterize
microscopically because of the larger number of possible terms in the binding
free energy. Explicit measurement
of the in-plane orientation would be useful~\cite{gomez2016}, for example using
electron microscopy~\cite{davies2012} or polarized
light~\cite{rosenberg2005}. Finally, the present study is limited to sensing
pure curvature, and it might be interesting to extend this analysis to consider
curvature gradients as well, since they have different symmetry properties
\cite{elias-wolff2018}.

%% file: curvsense_main.bbl
\providecommand{\noopsort}[1]{}\providecommand{\singleletter}[1]{#1}%

%% file: methods.ttt
\section{Supplementary methods}\label{sec:SImethods}

\paragraph{Coarse-grained lipid model.}

We used the \citet{cooke2005pre} lipid model in our simulations. Lipid
molecules are represented by one head- and two tail beads (see
Fig.~\ref{fig:cgmodels}). The solvent is modeled implicitly, and lipid bilayers
are stabilized by long-range attractive interactions between tail beads. The
model is expressed in terms of the three scales $\sigma$ (length), $\epsilon$
(energy), and $\tau$ (time). As in our previous work \cite{elias-wolff2018}, we
choose tuning parameters $w_c=1.6\sigma$ (decay length of tail attraction) and
temperature $k_BT=1.08\epsilon$, to yield stable bilayers in the fluid phase
\cite{cooke2005pre}. Simulations were carried out with Gromacs 2016.3, 2016.5,
and 2018.1, at constant volume, using a leap-frog stochastic dynamics
integrator with time step $0.01\tau$ and friction constant $\Gamma=\tau^{-1}$.
Under these conditions, comparisons of lipid bilayer thickness and lipid
diffusion indicate $\sigma\approx 1$ nm and $\tau\approx 10$ ns
\cite{cooke2005}. Coordinates were saved every $500\tau$.

We constructed protein models by assembling an asymmetric transmembrane monomer
comprised of 54 tail-beads sandwiched between two layers of 9 head beads
(Fig.~\ref{fig:cgmodels}c). The monomer was given a wedged cross-section by
compressing the beads on one side, corresponding to curvature radii of
$5\sigma$ and $25\sigma$, in the transverse and longitudinal directions,
respectively. We then formed multimeric protein structures by combining
monomers rotated around the membrane normal (Fig.~\ref{fig:cgmodels}d-g), and
stabilized the resulting structures by elastic springs of strength
$100\epsilon/\sigma^2$ between all beads closer than $4\sigma$. We have
previously studied the trimer in this family, and found a preferred curvature
of about $0.035\sigma^{-1}$ \cite{elias-wolff2018}. We used custom Matlab
scripts based on the mxdrfile package~\cite{mxdrfile} to construct the elastic
network models, generate initial configurations, and to analyze data sets.

\begin{table}[b]
  \begin{tabular}{c|cccrrr}
    name&$M$&$R\,[\sigma]$&$H\,[\sigma]$&$N_\text{lip}$&
         $N_\text{1M}$&$N_\text{2M}$\\
    \hline
    C1R07 & 1 & 7 & 60 & 4550 & -   & 32 \\
    C1R10 & 1 & 10 & 60 & 6520 & -  & 27 \\
    C1R15 & 1 & 15 & 75 & 12290& 83 & 17 \\
    C1R20 & 1 & 20 & 75 & 16420& 75 & 25 \\
    C1R25 & 1 & 25 & 75 & 20480& 75 & 25 \\
    C1R33 & 1 & 33 & 75 & 27060& 25 & 75 \\
    C1R50 & 1 & 50 & 75 & 41060&  - & 34 \\
    \hline
    C2R20 & 2 & 20 & 75 & 16395& 25 & - \\
    \hline
    C3R05 & 3 & 5 & 60 & 3225  & - & 25 \\
    C3R07 & 3 & 7 & 60 & 4515  & - & 25 \\
    C3R10 & 3 & 10 & 60 & 6475 & - & 44 \\
    C3R15 & 3 & 15 & 75 & 12240& - & 100 \\
    C3R20 & 3 & 20 & 75 & 16370& - & 100 \\
    \hline
    C4R10 & 4 & 10 & 60 & 6460 & - & 50\\
    C4R15 & 4 & 15 & 75 & 12230& - & 54\\
    C4R20 & 4 & 20 & 75 & 16355& - & 55 \\
    C4R25 & 4 & 25 & 75 & 20420& - & 54\\
    C4R33 & 4 & 33 & 75 & 27000& - & 62\\
    C4R50 & 4 & 50 & 75 & 41000& - & 26 \\
    \hline
    C5R20 & 5 & 20 & 75 & 16345& - & 25 \\
    \hline
    C6R20 & 6 & 20 & 75 & 16330& - & 25 \\
    \hline
  \end{tabular}
  \caption{Summary of production runs. $M$: number of protein subunits, $R$:
    cylinder radius, $H$: cylinder height, $N_\text{lip}$: approximate number of
    lipids (varies by less than $\pm20$ between same name replicas) ,
    $N_\text{1M}$: number of $10^6\tau$ production runs, $N_\text{2M}$: number
    of $2\times 10^6\tau$ production runs.}
  \label{SI:production}
\end{table}

\paragraph{Simulating curvature sensing on cylinders.}
For curvature sensing simulations, we assembled cylindrical bilayers with a
mid-monolayer density close to that of a flat bilayer. A protein molecule was
inserted in the bilayer and clashing lipids were removed. We assembled multiple
replicas for every combination of multimer and cylinder size, distinguished by
different initial protein in-plane orientation and by small random
long-wavelength perturbations to all particles. After energy minimization and a
burn-in simulation of $10^3\tau$, we ran production runs for $10^6\tau$ or
$2\times10^6\tau$, saving positions every $500\tau$. Relaxation of cylindrical
systems is facilitated by the artificially high flip-flop rate of the Cooke
model~\cite{harmandaris2006}. System size parameters and the number of
production runs are summarized in Table~\ref{SI:production}.

\paragraph{Data analysis.}
At every sample point, we computed the in-plane orientation by least-square
fitting straight lines through a selected set of core particles in each
monomer. The line was projected on the cylinder tangent plane intersecting the
protein midpoint, and the in-plane orientation of that subunit was computed as
the angle between the projected line and the $z$-axis.

Only data from subunit one is included in orientation angle histograms (e.g.,
Fig.~\ref{fig:cylhist}, left). For analyzing the free energy profiles and their
amplitudes (Fig.~\ref{fig:cylhist}, right), we included the in-plane
orientation of all subunits, as well as replicas rotated by $\pi$, to enforce
the 2-fold rotation symmetry of the curvature tensor itself. (That is, $2m$
values from each configuration of an $m$-mer.) We then aggregated data from all
simulations of each system size into a single histogram with number of bins
proportional to the protein multiplicity, and computed the free energy as the
negative log density. To compute the amplitude of the resulting free energy
profile, we first fitted a truncated Fourier series with the two lowest
wave-lengths, and then extracted the amplitude of the fit. Examples of such
fits are shown in red in Fig.~\ref{fig:cylhist}(right), displaying excellent
agreement.
Error bars in Figs.~\ref{fig:cylhist} and \ref{fig:amplitude} correspond to
95\% bootstrapped confidence intervals from $10^4$ bootstrap realizations.
These were computed from the aggregated data, taking only uncorrelated data
points (autocorrelation times range from $1200 \tau$ for the monomer, to
$8000 \tau$ for the dimer). The error bars in Fig.~\ref{fig:amplitude}b
correspond to relative errors, that is, for a given amplitude $y$ with bootstrap confidence intervals
$\pm \delta y$ in linear space, the relative error in logarithmic scale,
where $z = \log_{10}y$, corresponds
to $\delta z = (\log 10)^{-1} \delta y / y $.



%


%% file: curvsense_main.bbl
\begin{thebibliography}{32}
\bibitem [{num()}]{numterms}%
  \BibitemOpen
  \href@noop {} {}\bibinfo {note} {Each term of order $m$ in \Eq{eq:taylor} can
  be written on the form $h_1^{i}h_2^{j-i}h_3^{m-j}$, with $0\le i\le j\le m$.
  The number of distinct terms therefore equals the number of
  $i,j$-combinations, given by $\sum_{i=0}^m\sum_{j=i}^m 1=(m+1)(\frac
  m2+1)$.}\BibitemShut {Stop}%
\end{thebibliography}
